\documentclass[a4paper, 10pt]{article}
\usepackage[a4paper]{geometry}
\usepackage{hyperref}
\usepackage[english]{babel}
\usepackage[utf8]{inputenc}
\usepackage{graphicx}
\usepackage{listings}
\usepackage{verbatim}
\usepackage{amsmath}
\usepackage{cite}
\usepackage{placeins}
\usepackage{url}
\usepackage{float}
\usepackage{titling}
\usepackage{multirow}
\usepackage{xcolor}
\usepackage[affil-it]{authblk}
\usepackage[font={footnotesize}]{caption}
\newcommand{\e}{\text{e}}

\begin{document}

\title{Improving wastewater-based epidemiology through strategic placement of samplers}
\author[1]{A. J. Wood}
\author[2]{J. Enright}
\author[1]{A. Sanchez}
\author[3]{E. Colman}
\author[1,4]{R. R. Kao}
\affil[1]{Roslin Institute, University of Edinburgh}
\affil[2]{School of Computing Science, University of Glasgow}
\affil[3]{Bristol Medical School, University of Bristol}
\affil[4]{Royal (Dick) School of Veterinary Studies, University of Edinburgh}
\maketitle


\begin{abstract}
Wastewater-based epidemiology (WBE) is a fast emerging method for passively monitoring diseases in a population. By measuring the concentrations of pathogenic materials in wastewater, WBE negates demographic biases in clinical testing and healthcare demand, and may act as a leading indicator of disease incidence. 

For a WBE system to be effective, it should detect the presence of a new pathogen of concern early enough and with enough precision that it can still be localised and contained. In this study, then, we show how multiple wastewater sensors can be strategically placed across a wastewater system, to detect the presence of disease faster than if sampling was done at the wastewater treatment plant only. Our approach generalises to any tree-like network and takes into account the structure of the network and how the population is distributed over it.

We show how placing sensors further upstream from the treatment plant improves detection sensitivity and can inform how an outbreak is evolving in different geographical regions. However, this improvement diminishes once individual-level shedding is modelled as highly dispersed. With overdispersed shedding, we show using real COVID-19 cases in Scotland that broad trends in disease incidence (i.e.,~whether the epidemic is in growth or decline) can still be reasonably estimated from the wastewater signal once incidence exceeds about 5 infections per day.
\end{abstract}


\section{Introduction}
Wastewater-based epidemiology (WBE) is an emerging method for the passive monitoring of diseases in a population. In WBE wastewater material is sampled from the sewer network, and the concentration of one or more biomarkers (such as fragments of pathogenic material) is measured~\cite{sims2020future,gracia2017measuring,boogaerts2021current,kilaru2023wastewater}. After taking into account factors such as the population living upstream from the sample location and the amount to which the wastewater sample has been diluted by other sources such as rainwater and industrial waste, the concentration of pathogenic material can inform what the prevalence of disease in a population may be, and how it is changing. A substantial advantage of WBE is that it allows diseases (including asymptomatic infections) to be monitored passively over a whole population without the need for potentially costly and manual clinical testing, and also negates demographic biases in testing and reporting. Depending on the shedding profile (the amount of shed material over time after infection), WBE may serve as a leading indicator of disease incidence over clinical cases, or healthcare demand~\cite{colman2025impact}.

WBE as a monitoring tool expanded dramatically during the COVID-19 pandemic~\cite{fitzgerald2021site,morvan2022analysis,fang2022wastewater}. For WBE to become a generalisable, long-term monitoring method for diseases such as COVID-19 a key challenge is in understanding how a measured concentration of pathogenic material maps to a real-world disease prevalence, taking into account factors such as variation in shedding behaviour, dilutants and biomarker degradation~\cite{wade2022understanding}. 

The drainage area of a particular treatment plant may comprise a sprawling network of connected pipes. The flow of wastewater towards the treatment plant can be thought of as a tree network, with all edges directed towards a single root node. With some minor exceptions (see Fitzgerald, Rossi et al.~\cite{fitzgerald2021site} for a description of the system), all wastewater sources within a geographically defined area, can be assumed to contribute to this system with source points all the way up the tree's branches.

The concentration of a biomarker in the wastewater will vary depending on which node the sample is taken from. In an ideal scenario, a monitoring sensor at the treatment plant at the treatment plant should capture all pathogenic material and give the concentration over the whole population, but does not provide any geographical information below the drainage area level, and may suffer from poor detection when there are few infected individuals contributing to the system. This is of particular concern for diseases where any presence at all would be of great concern e.g.~polio in the UK~\cite{klapsa2022sustained}. Subsampling higher up the system may improve these results, but suffer from higher levels of noise, and may add unwanted or unnecessary burdens on sampling and laboratory capacity.

Recognising the profound real-world challenge of sample normalisation in WBE, we use this study to focus on the structure of the wastewater networks themselves and ask: when setting up a WBE system to effectively detect new diseases early and monitor longer-term trends with a degree of spatial resolution, then, where should sensors be placed? We treat this as a network problem, and explore a strategy for sensor placement in a general tree network. Second, placing sensors upstream results in lower populations (and likely infected populations) residing upstream from each sensor. The amount of pathogenic material shed at the individual-level is likely to be highly overdispersed~\cite{arts2023longitudinal}. Using simulation models we assess how such differences in shedding lead to variability in the wastewater signal over time, and examine the relationship between disease incidence, and the quality of the resulting wastewater signal in terms of indicating whether an epidemic is growing or shrinking in size.

We use data on wastewater networks in Scotland, and analyse wastewater networks of three catchment sizes: small ($\sim 20\,000$ residents), medium ($\sim 190\,000$) and large ($\sim 580\,000$). For disease incidence, we use simulated outbreaks, as well as real COVID-19 PCR positve tests in Scotland from the years 2020--2022.


\section{Methods and data}

\subsection{Sensor placement}
We treat a wastewater system as a rooted tree with all edges directed towards a single root node (the wastewater treatment plant). Each node has an associated resident population.

For $M$ sensors $m = (1, 2, 3, \dots, M)$, we define a sensor's catchment as the population residing upstream from it, whose wastewater will pass through that sensor's location. Wastewater may pass through multiple sensors before reaching the treatment plant, thus in multi-sensor systems many individuals may fall into more than one catchment. We say that two nodes in the tree are indistinguishable if they are upstream of exactly the same set of sensors: that is, if a disease signal originating in those nodes would be detected by exactly the same set of sensors.  Therefore a signal detected at that set of sensors cannot be pinpointed to within any set of indistinguishable nodes with respect to those sensors.  Where we have population estimates at each node, we weight nodes by that population when calculating the size of an indistinguishable set.  

In a directed rooted tree with a single sink, the problem of placing sensors so as to minimise the maximum size of an indistinguishable set is equivalent to placing sensors so that removing the edges from the nodes where the sensors are places to their downstream parent node minimises the maximum size of a remaining component in the network.    

Our placement algorithm uses a dynamic programming approach to essentially iterate over all possible optimal placements of sensors with respect to this maximum component size measure, and is directly adapted from the approach used in Reference~\cite{enright2018deleting},  which gives an exact solution to this problem on trees. As the algorithm seeks sensor placements with respect to the maximum component size and not a specific number of sensors, solutions are not returned for all possible numbers of sensors.

\subsection{Data}
\subsubsection*{Wastewater}
High-resolution sewer pipe shapefile data are provided by the Scottish Environment Protection Agency (SEPA), broken down by drainage operational area (DOA). We focus on Edinburgh (DOA population $\sim 580\,000$), Meadowhead ($\sim 190\,000$), and Peterhead ($\sim 20\,000$). Full details of tree construction are provided in Supplementary Material~\ref{sec:treebuilding}. Briefly here, using the $\texttt{sf}$ and $\texttt{igraph}$ packages in R we convert the sewer pipes into a list of edges, and the connections between those pipes as a list of nodes. We remove cycles and assume that all wastewater takes the shortest available route, by physical distance, to the treatment plant, and we do not differentiate between surface water/foul/industrial waste pipes. We then assign the population to nodes using the 2022 census data at the level of Output Area (OA, geographical census areas, each containing $\sim 100$ residents)~\cite{Census2022Population}. The final trees likely simplify the real passage of wastewater, but should capture the fundamental structure of the wastewater network and how the human population structure projects onto it.

\begin{figure}
	\includegraphics{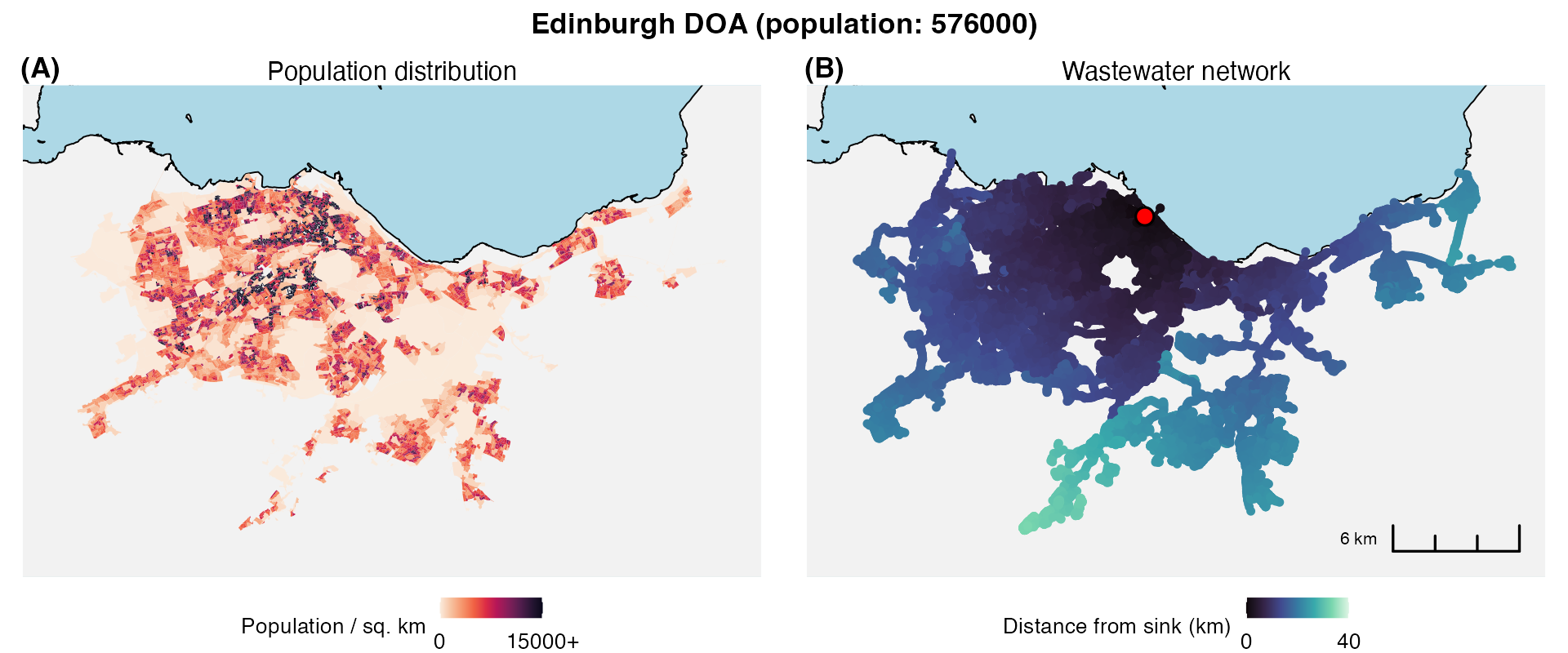}
	\caption{The Edinburgh drainage operational area (DOA). (A) Population distribution. (B) the corresponding wastewater network, where the colour indicates the distance from the treatment plant (marked with a red dot).}
	\label{fig:E_population_and_network}
\end{figure}

Figure~\ref{fig:E_population_and_network} shows the population distribution and wastewater tree for the Edinburgh DOA (see Supplementary Material, Figures~\ref{fig:M_population_and_network},~\ref{fig:P_population_and_network} for Meadowhead and Peterhead respectively).

\subsubsection*{COVID-19 cases}
COVID-19 case data are provided from Public Health Scotland's \emph{electronic Data Research and Innovation Service} (eDRIS) system. The data include individual polymerase chain reaction (PCR) as well as lateral flow device (LFD) tests, with test result, age, sex, and residing data zone (a census area typically comprising $500$--$1\,000$ individuals). De-identified IDs link repeat tests by the same individual. These metadata --- in particular the DZ, specifying location to within an area as small as 0.1 km$^2$ in densely populated areas --- therefore identify cases at a fine spatio-temporal scale.

\subsection{Modelling individual-level shedding}
Precise individual-level data on pathogenic shedding and changes over the course of infection are limited. For COVID-19, authors in Reference~\cite{arts2023longitudinal} monitored the presence and concentration of the SARS-CoV-2 N gene in stool samples from 48 infected individuals, up to 28 days after symptom onset. Positive samples were recorded from 6 out of 7 stool samples taken 2 days after symptom onset. The size and longitudinal profile of measursed concentration is strongly overdispered; from the authors' raw data in Reference~\cite{arts2023longitudinal}, Supporting Information Figure S2, the logarithm of the peak N gene concentration from samples that exceeded their detection threshold had a standard deviation of $3.07$.

To reflect shedding overdispersion in our analysis, for each infected individual $i$'s shedding $s_i(t)$, we sample a peak shedding, $s^*_i$ from a lognormal distribution; $\log{s_i}$ is normally distributed with some mean value $\mu$ and variance $\sigma^2>0$. $s_i$ therefore has a mean value of $\langle s_i \rangle = \e^{\,\mu + \frac{1}{2}\sigma^2}$, median $\e^{\,\mu} < \langle s_i \rangle$. We fix $\mu=0$ (so the median peak shedding is $\e^0 = 1$), and $\sigma = 3.07$. We take shedding as increasing from zero linearly to this maximum $s_i^*$ over the course of 5 days after infection (in line with the estimated incubation period of SARS-CoV-2~\cite{lauer2020incubation}). The avaiable data are limited for estimating temporal shedding but from the study in Reference~\cite{arts2023longitudinal} of samples taken 28 days after symptom onset, 2 out of 20 still returned a positive sample. As a tentative interpretation, we then take shedding to decrease linearly over 23 days, so an individual sheds for 28 days total.

The concentration recorded at sensor $m$ at time $t$ is the aggregated signal from all individuals upstream from $m$ divided by the total population $N_m$:
\begin{align}\label{eq:aggregation}
	S_m(t) = \sum_{\substack{\text{Infectees } i \text{ in} \\ \text{catchment of } m}}  \frac{s_i(t)}{N_m}
\end{align}
 Due to how sensors are placed an individual may reside in the catchments of two or more sensors. We impute a detection threshold of 1 in $10\,000$ as estimated for SARS-CoV-2 in Reference~\cite{hewitt2022sensitivity}, with a non-zero signal registered in an upstream population of $10\,000$ if the concentration exceeds that of one individual shedding at the median peak shedding. There is no delay between shedding and detection at a sensor, and no additional signal noise is modelled.

\subsection{Time to detect novel outbreaks}
To estimate how quickly new outbreaks can be detected under different sensor configurations, we simulate 200 disease outbreaks over the DOA's residing population. For each we sample a simple exponential incidence trajectory with a doubling time between 2 and 6 days. We randomly assign each infection event $i$ to a resident (and their residing node), sample a shedding profile $s_i(t)$, and aggregate signal per Eq.~\ref{eq:aggregation}. The detection speed is the cumulative disease incidence across the whole DOA at the time the signal at any of the sensors exceeds the detection threshold.

For overdispersed shedding, we evaluate detection speed for configurations of up to 20 sensors, where samples are taken either daily, every three days, or every seven days. As a point of comparison we do the same analyses for a scenario where all individuals have an identical shedding curve, with a peak of 1 and the same 28-day temporal profile.

\subsection{Relationship between incidence and wastewater signal}
For longer-term monitoring, we want to assess the relationship between disease incidence $I(t)$ and the wastewater signal $S(t)$, and how an overdispersed shedding distribution affects the quality of the wastewater signal. Specifically, we want to quantify at what level of community incidence the wastewater signal can reasonably inform us about the growth rate (positive or negative) of disease incidence. As $S(t)$ and $I(t)$ are measured on different scales but each typified by phases of exponential growth and decay, comparing their logarithms here is more appropriate than raw magnitudes. This linearises the dynamics and allows an interpretable comparison of estimates of epidemic growth and decline, rather than focusing on raw magnitudes.

For each DOA we take COVID-19 case data from September 2020 to August 2022. We again associate to each case to an individual $i$ (now based on the data zone the case was originally reported from), and sample a shedding profile $s_i(t)$. For each sensor $m$, we count incidence within the catchment $I_m(t)$, and aggregate shedding to find a wastewater profile $S_m(t)$. For simplicity we assume for this part of the analysis that sensors have perfect sensitivity (i.e.~that incidence is sufficiently high that a signal can be consistently detected).

To establish a ground truth of epidemic growth to compare the wastewater signal against, using a method described by Colman and Kao in Reference~\cite{colman2025impact}, we fit to $I_m(t)$ a smoothed trajectory $I_m^*(t)$, through a series of connected exponential curves:
\begin{align*}
	I_m^*(t) = 
	\begin{cases}	
		\e^{\lambda_1 t} - 1& \text{if } 0\leq t < t_1\\
		\left(\e^{\lambda_1t_1} -1\right) (\e^{\lambda_2(t-t_1)}) & \text{if } t_1 \leq t < t_2 \\
		\left(\e^{\lambda_1t_1} -1\right)(\e^{\lambda_2(t_2-t_1)})(\e^{\lambda_2(t-t_2)}) & \text{if } t_2 \leq t < t_3\\
		\vdots 
	\end{cases}	\;,
\end{align*}
fitting the inflection points $(t_1, t_2, t_3, \dots)$, and the growth rates $(\lambda_1, \lambda_2, \lambda_3, \dots)$ (which can take positive or negative values if the epidemic is in growth or decline respectively). The log of $I_m^*(t)$ has gradient equal to the growth rate at time $t$:
\begin{align*}
	\log{I_m^*(t)} \propto 
	\begin{cases}	
		\lambda_1 t  & \text{if } 0\leq t < t_1\\
		\lambda_2 t & \text{if } t_1 \leq t < t_2 \\
		\lambda_3 t & \text{if } t_2 \leq t < t_3\\
		\vdots 
	\end{cases}	\;.
\end{align*}
The epidemic growth rate estimated from the wastewater signal is then the gradient of $\log{S_m(t)}$. Comparing to the ground truth over some interval $[t_a, t_b]$, define the error $\epsilon(t)$
\begin{align*}
	\epsilon(t) = \left[\log{I^*_m(t)} -  \log{I^*_m(t_a)}\right] - \left[\log{S_m(t)} - \log{S_m(t_a)}\right] \;.
\end{align*}
By aligning the log signals at $t = t_a$ and fixing $\epsilon(0) = 0$, $\epsilon(t)$ tracks how far the estimated growth rate deviates from ground truth growth rate, over that time interval. Finally, define the signal-to-noise ratio (SNR) as
\begin{align*}
	\text{SNR} = \frac{\sum_{t = t_A}^{t_B} \log^2{I^*_m(t)}}{\sum_{t=t_A}^{t_B}\epsilon(t)^2}\;.
\end{align*}
An SNR above 1 indicates that the power of $\epsilon(t)$ is lower than the power of the signal $\log{I_m^*(t)}$ being estimated, thus the true growth rate can be more clearly distinguished in $\log{S(t)}$ against the background of fluctuations introduced by individual variation in shedding.

To assess the relationship between SNR and disease incidence we compare the SNR over a 28-day rolling window to the mean disease incidence in the same period, across all sensors and DOAs.


\section{Results}

\subsection{Sensor placement}
\begin{figure}
	\centering
	\includegraphics[trim = 0cm 0.7cm 0cm 0.8cm, clip]{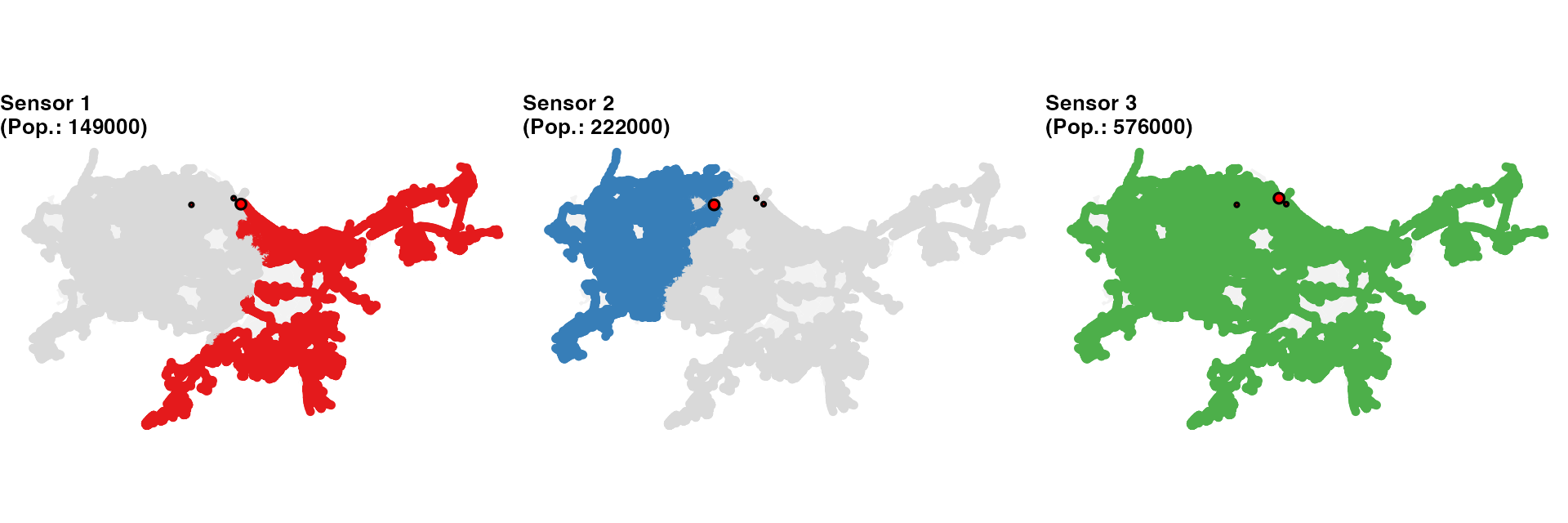}
	\rule{\textwidth}{0.5pt}
	\includegraphics[trim = 0cm 1.4cm 0cm 1.5cm, clip]{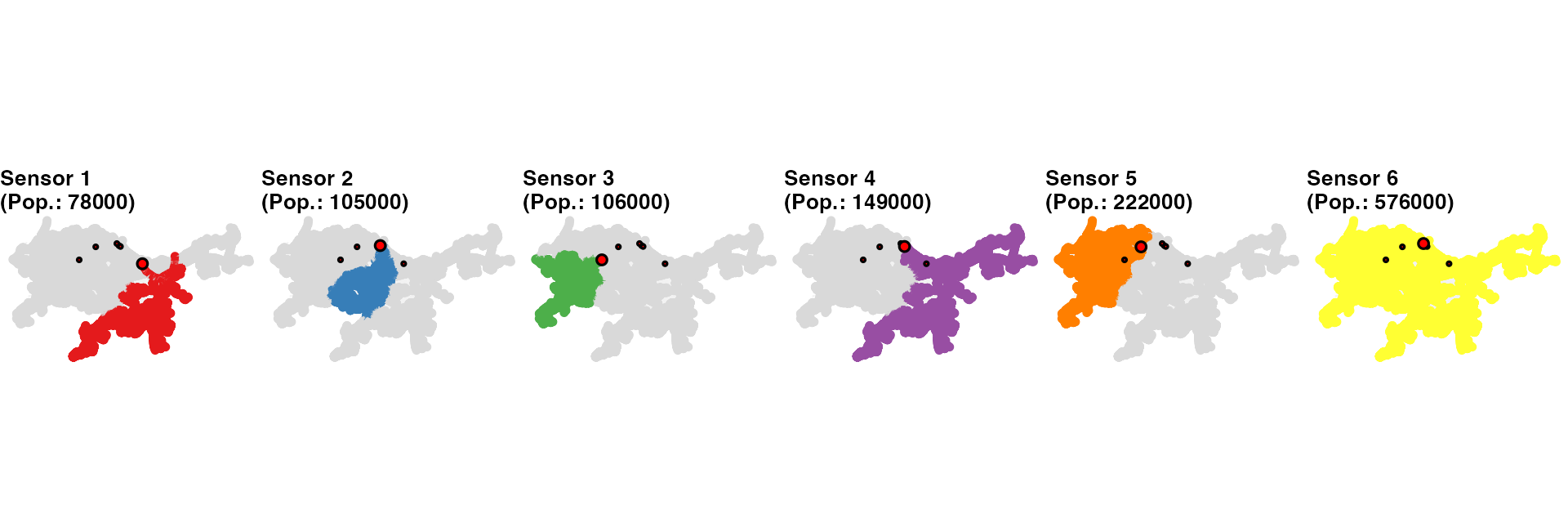}
	\rule{\textwidth}{0.5pt}
	\includegraphics[trim = 0cm 1.5cm 0cm 1.5cm, clip]{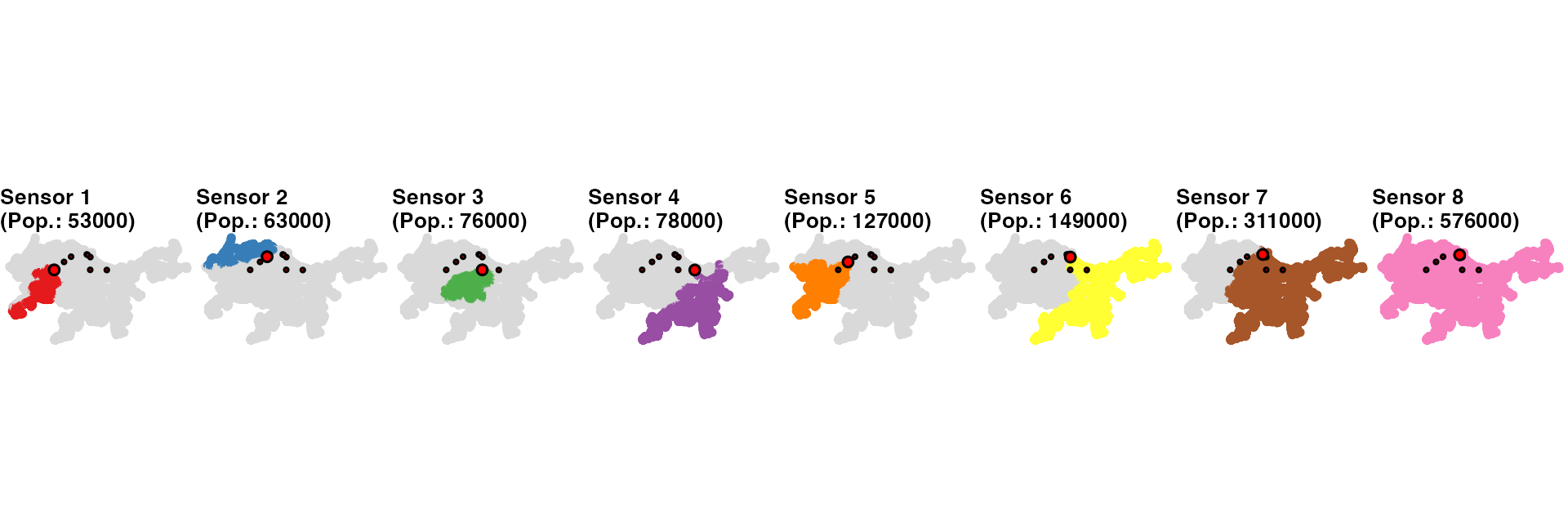}
	\caption{Placement of sensors over the Edinburgh DOA, for (top to bottom) 3, 6, and 8 sensors.}
	\label{fig:E_sensor_placement}
\end{figure}

Figure~\ref{fig:E_sensor_placement} shows the sensor placement and corresponding catchments for the Edinburgh DOA (equivalent for the smaller Meadowhead and Peterhead DOAs in Supplementary Material, Figures~\ref{fig:M_sensor_placement},~\ref{fig:P_sensor_placement}). Each setup has a a single sensor placed at the treatment plant (required to ensure the whole population is upstream from at least one sensor), and other sensors capture different regions of the catchment area, with a degree of overlapping, especially as the number of sensors increases.

\subsection{Relationship between incidence and wastewater signal} 
\begin{figure}
	\includegraphics[width=0.99\textwidth]{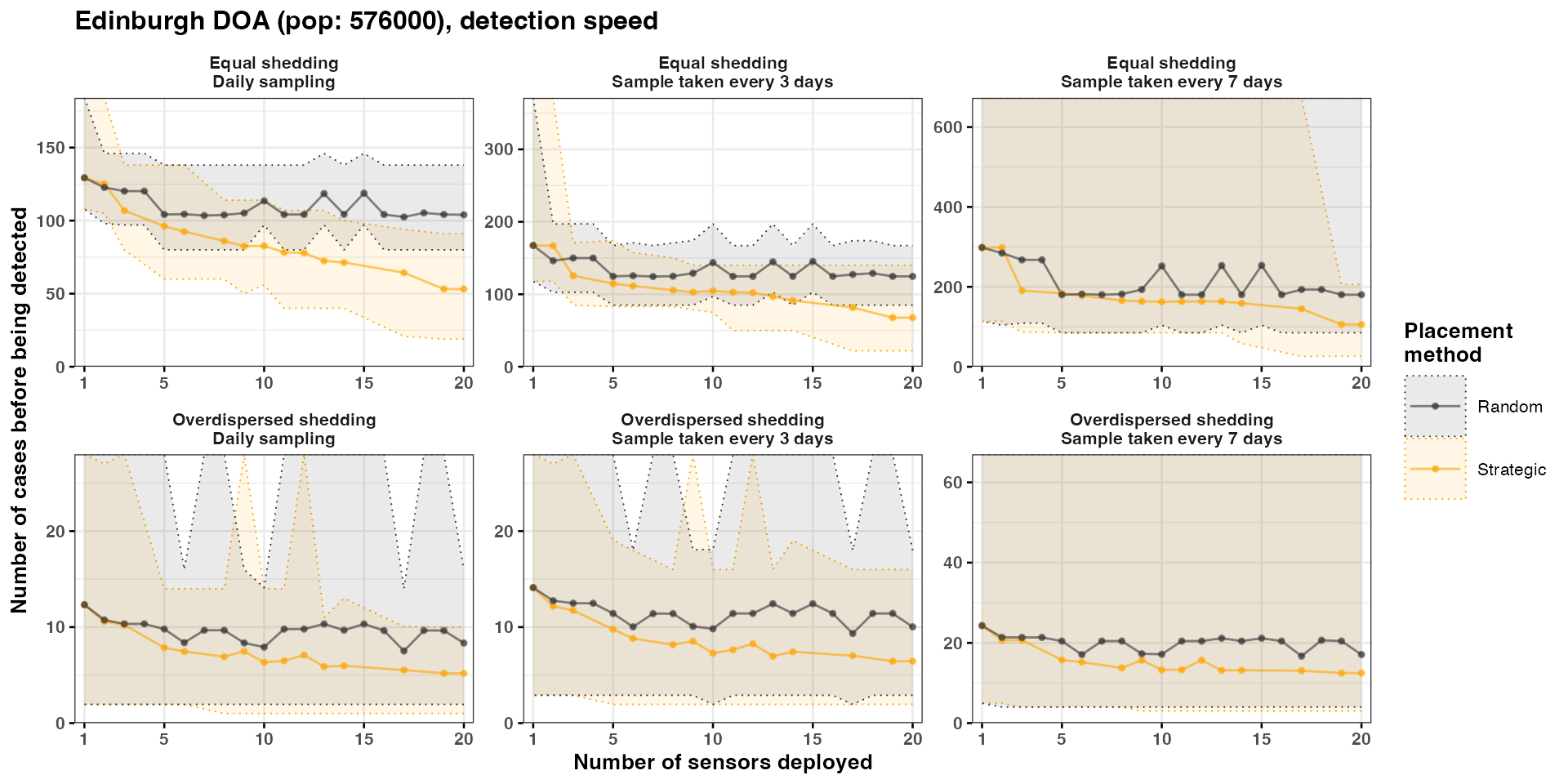}
	\caption{Detection speed for different sensor configurations, in the Edinburgh DOA. Detection speed is quantified by the cumulative disease incidence by the time any sensor detects a non-zero signal. Top: results when shedding is equal (all infected individuals have the same shedding profile with a mean of 1). Bottom: results when shedding is overdispersed (peak shedding is sampled from a lognormal distribution with log mean 1, log standard deviation 3.07).}
	\label{fig:E_first_detection}
\end{figure}

Figure~\ref{fig:E_first_detection} compares detection speed and variation (and 90\% confidence interval (CI)) to sensor placement strategy, number of sensors deployed, sampling frequency and shedding pattern, for the Edinburgh DOA. For a single-sensor setup with daily sampling, the mean cumulative incidence at the time of detection is 129 [108, 184] when shedding is equal across all individuals, as opposed to 12.3 [2, 28] when shedding is overdispersed. This is an instance of a more general trend where outbreaks are detected earlier when shedding is overdispersed. On increasing the number of sensors deployed strategically, the mean cumulative incidence at detection falls. For equal shedding, a ten-sensor setup detected outbreaks after a mean cumulative incidence of 82.7 [56, 114] (c.f.: 113 [97, 138] with ten randomly placed sensors), and for twenty sensors improved further to 53.2 [19, 91] (c.f.:~ 104 [80, 138]). However, this relative improvement is diminished when shedding is overdispersed; with a smaller comparative improvement on increasing the number of sensors, and also comparing strategic placement to random placement. Finally, Figure~\ref{fig:E_first_detection} shows detection speed falls markedly across all setups, once sampling frequency decreases from once per day. For infrequent sampling taken every 7 days, the 90\% CI is markedly wider. 

These same patterns are broadly captured for the smaller Meadowhead and Peterhead DOAs (see Supplementary Material~\ref{supp:bonusfigs}, Figures~\ref{fig:M_sensor_placement},~\ref{fig:P_sensor_placement}). Both DOAs have smaller populations and in turn outbreaks are generally detected at a lower cumulative incidence.

\subsection{Effect of catchment area size on signal quality}
\begin{figure}
	\includegraphics[width=0.99\textwidth]{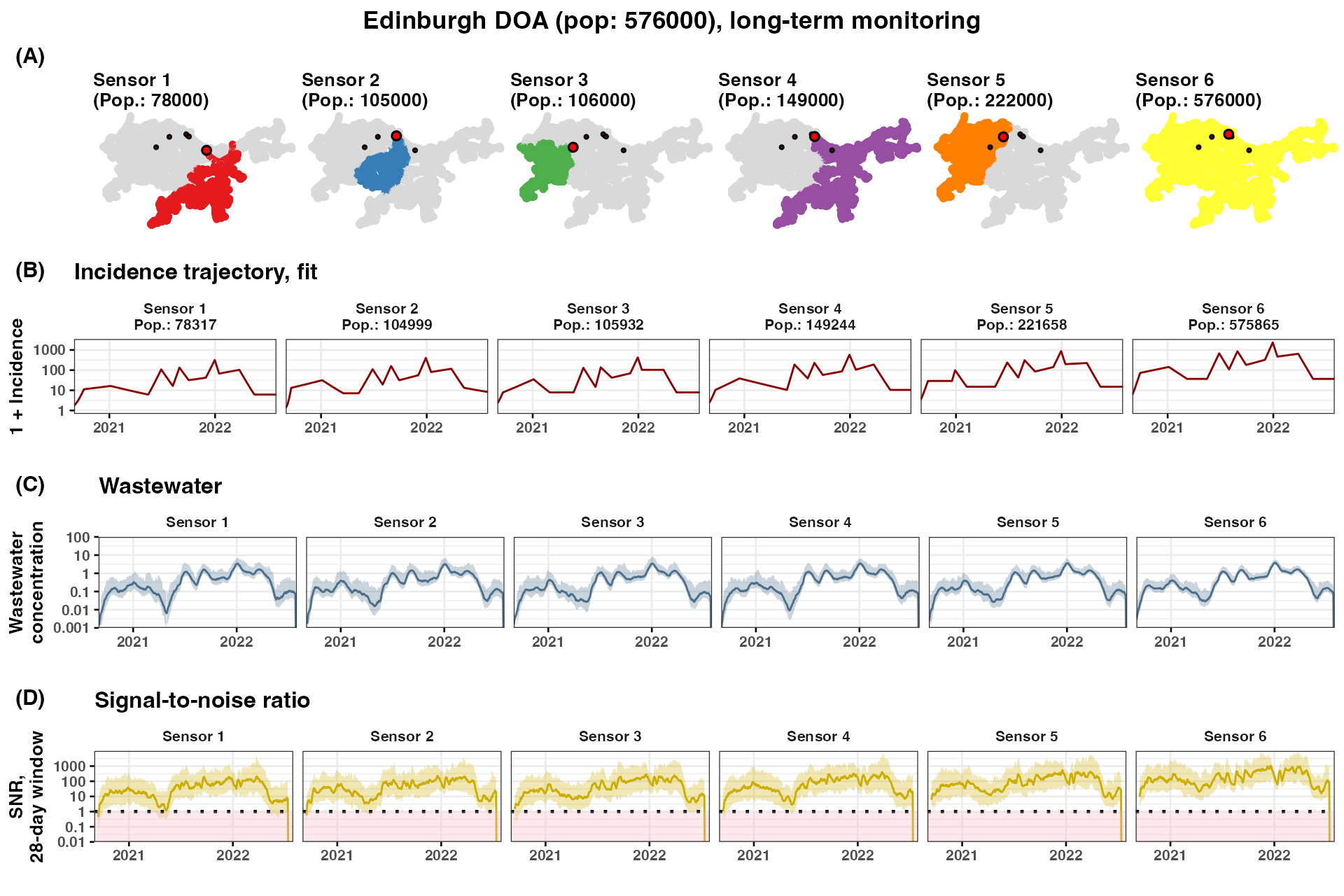}
	\caption{Simulated long-term monitoring of disease in the Edinburgh DOA via wastewater. (A) Placement of an 6-sensor configuration, with sensors ordered by population. (B) Smoothed COVID-19 incidence trajectory between September 2020 and August 2022. (C) Modelled wastewater signal with the line indicating the median signal, and the filled region the [5\%, 95\%] range of signals over different shedding intensity samples. (D) The signal-to-noise ratio of the wastewater signal relative to the smoothed incidence trajectory, over a 28-day rolling window.}
	\label{fig:E_longterm}
\end{figure}

Figure~\ref{fig:E_longterm} shows simulated wastewater surveillance for an six-sensor setup in the Edinburgh DOA, with real-world disease incidence, smoothed incidence, modelled wastewater signal and the evaluated SNR. This shows variation (noting the log scale) in the wastewater signal across different simulations, each with different samples from the overdispersed shedding distribution. The signal-to-noise measure is characteristically higher and more stable in periods of higher incidence, indicating the growth rate implied from the wastewater signal more closely tracks the ground truth from disease incidence. Equivalent plots for the Meadowhead and Peterhead DOAs are presented in Supplementary Material~\ref{supp:bonusfigs}, Figures~\ref{fig:M_longterm},~\ref{fig:P_longterm} respectively.

\begin{figure}
	\includegraphics[width=0.99\textwidth]{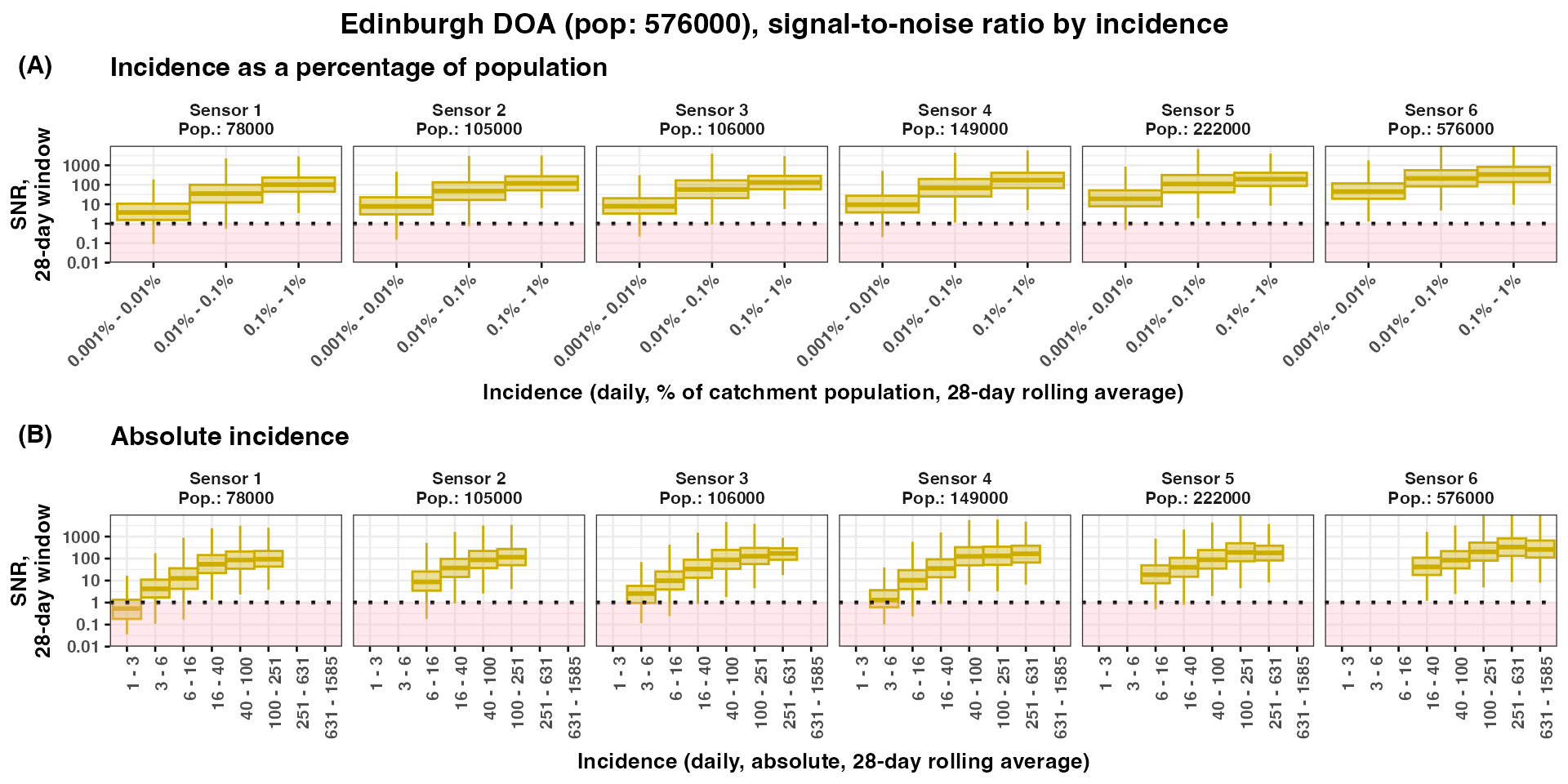}
	\caption{Signal-to-noise ratio statified by disease incidence in the Edinburgh DOA, with 6 sensors. (A) Incidence as a percentage of the catchment population. (B) Incidence as an absolute number of infections.}
	\label{fig:E_longterm_SNR}
\end{figure}

Figure~\ref{fig:E_longterm_SNR} shows SNR stratified against disease incidence, showing that SNR generally exceeds 1 once daily incidence exceeds about 3--6 infections within a sensor's catchment. Equivalent plots for the Meadowhead and Peterhead DOAs are presented in Supplementary Material~\ref{supp:bonusfigs}, Figures~\ref{fig:M_longterm_SNR},~\ref{fig:P_longterm_SNR} respectively, showing that this SNR-to-incidence relationship remains robust despite them being different drainage areas and outbreaks.


\section{Discussion}
This work shows how strategic placement of sensors upstream in a sewage system can result in more effective monitoring of a disease in large populations, as compared to a default method of taking samples at the wastewater treatment plant only, where all wastewater across the whole population has aggregated. We have shown a strategy for dividing a large wastewater system into smaller population groups, which in principle could be monitored independently in the long-term. Our approach is applicable to any acyclic wastewater system that has a single treatment plant at its root, that all sewer pipes flow towards.

Placing multiple sensors further upstream alleviates the real-world problem of sensors having an imperfect sensitivity; taking the Edinburgh DOA (population $\sim 580\,000$) as an example, a sensitivity of 1 in $10\,000$ (i.e.~can potentially detect as few as one infected individual in a catchment of $10\,000$) means that at least $\sim$60--100 infections may have already occurred by the time a single sensor at the treatment plant could detect anything (and more if those infections are drawn out in time), which depending on the disease may already be unfeasible to contain. Depending on what prevalence of a disease may be deemed acceptable before it is detected, our results provide estimates of how many sensors should be deployed, for catchment areas of different sizes.

\subsubsection*{Limitations}
In this work the modelled disease outbreak, individual shedding, wastewater signal and sampling are idealised.

On shedding profiles and sensor sensitivity, while we include a concentration threshold for detection in our sampling, sampling in this model is instant, the measured concentration once signal exceeds the detection threshold is perfect, and variation between different simulations of the same incidence trajectory is introduced from individual-level variation in shedding only. While we have included high overdispersion in individual-level shedding, we assume all individuals shed for the same duration with the same profile. We expect the tail of the shedding profile to be less important when estimating time-to-first-detection; in a growing epidemic, recent infections situated at the beginning at the shedding curve will dominate numerically. 

We assume that infected individuals shed at their place of residence only, which ignores time spent at e.g.~work or school, and also individuals that e.g.~work in the DOA but do not live there. Also, while less important in a presence/absence system (where any signal at all is concerning enough to public health officials), the translation from wastewater signal measurement to disease prevalence is a substantial problem at both the biological scale (how much pathogenic material an infected person sheds into the wastewater) and the system scale (how much wastewater, rainwater and industrial waste is flowing through the sampler).

Notwithstanding these additional real-world challenges, this work still makes an important contribution in how more comprehensive future WBE-based monitoring systems should be strategically designed around the structure of the population and underlying sewer network, especially for presence-absence detection where the only trigger for control measures is a non-zero signal of any amount.

\subsubsection*{Speed of detection and overdispersed shedding}
Our method for sensor placement detects new outbreaks of disease significantly faster (that is, at a lower cumulative incidence) than placing the same number of sensors at random. By strategically reducing the population upstream from any given sensor, the mean number of cases required to exceed a detection threshold reduces, and a non-zero signal is registered earlier in an outbreak.

In the real-world, shedding is likely to be overdispersed. Infrequent ``super-shedders'' may shed orders of magnitude more pathogenic material than the median, thereby disproportionately influencing any aggregate signal. When we make shedding overdispersed, detection speed improves overall compared to when all individuals shed the same amount; while the median is fixed, the mean shedding is a factor $\e^{\sigma^2/2}$ (here: $\e^{3.07^2/2} \approx 111$) greater, thus the mean number of infections required to exceed a given detection threshold falls. However, the comparative improvement in detection speed offered by strategic placement of sensors is diminished. When shedding is overdispersed the time to detection becomes contingent on the infection times of those ``super-shedders'' that are more likely to clearly exceed a sensor's detection threshold, diminishing the relative importance of factors such as the sensor distribution and quantity. Nonetheless, the smaller average catchment achieved with strategic sensor placement still provides higher precision on where infected individuals are likely to be, over sampling at the treatment plant alone or placing sensors randomly.

A straightforward result is that for improving detection speed, that high sampling frequency is more important than the number of sensors, i.e.~a daily sample at the treatment plant will likely outperform sampling at two locations every two days, or three locations every three days etc. If a disease has a short doubling time, even a short time gap between samples may permit a substantial number of infections to occur between samples, and instead a single high-frequency sensor will likely detect the disease earlier and, in the instance of overdispersed shedding, decrease the probability of a ``super-shedder'' being missed entirely between samples.

\subsubsection*{Utility of wastewater signal for inferring epidemic growth and decline phases}
For longer-term monitoring, it is important to understand what changes in the wastewater signal tell us about changes in the epidemic. In the limit of large incidence the variation in the wastewater signal becomes normally distributed as moments higher than the standard deviation fall to zero~\cite{colman2025impact}. This is regardless of how overdispersed individual-level shedding is (though, of course, what this limit of large incidence is does depend on the magnitude of overdispersion). 

Here we have explored the utility of the wastewater signal in scenarios where incidence is lower and the signal is more skewed by ``super-shedders''. While disease incidence and the wastewater signal are quantified by different units, if the wastewater signal were a perfect proportionate analogue to disease incidence, the gradients of the logarithms of their timeseries, being the exponential growth rate, should be equal. Using a signal-to-noise -like measure, we have measured how closely the wastewater signal tracks the rate of epidemic growth estimated from fitting disease incidence, using real-world COVID-19 data. In our model, despite overdispersed shedding, once daily incidence exceeds of order 3--6 infected individuals, the SNR generally exceeds 1 and the true growth rate can be more clearly distinguished against the noise of background fluctuations. This threshold is consistent across the different DOAs explored.

Our interest here was in quantifying wastewater signal quality against absolute incidence, so for simplicity we assumed in our long-term modelling simulations that all sensors had perfect sensitivity. As the SNR is dependent upon the absolute number of infections, for low-incidence diseases the signal from larger catchment areas may be of more utility for characterising epidemic progression, aggregating more infections as compared to sensor further upstream, but only if shedding actually exceeds those sensors' detection threshold. Thus for long-term monitoring, sensors should be placed with the detection threshold and estimated incidence of the disease being monitored in mind; sampling closer to the sink sacrifices spatial resolution for likely higher signal quality, but should not be placed too downstream where the estimated incidence of disease is likely to fall below the detection threshold.

\section*{Acknowledgements}
We thank SEPA for the provision of wastewater data, and eDRIS for the provision of COVID-19 testing and severe outcomes data.

\section*{Funding statement}
This work has been funded by the ESRC grant ES/W001489/1.

\bibliographystyle{vancouver}
\bibliography{WW_references}

\clearpage

\clearpage

\setcounter{page}{1}

\section*{\LARGE{Supplementary Material}} 

\emph{\Large{Improving wastewater-based epidemiology through strategic placement of samplers}}

\section*{}

\clearpage

\appendix
\renewcommand\thefigure{S\arabic{figure}}    
\renewcommand\thetable{S\arabic{table}}    
\setcounter{figure}{0}  
\setcounter{table}{0}

\section{Constructing a tree from sewer shapefile data}\label{sec:treebuilding}
Analysis code is written in \emph{R} (version 4.3.1). We read and manipulate sewer shapefiles using the $\texttt{sf}$ (version 1.0.16), then use $\texttt{igraph}$ (version 2.1.4) to build and analyse the resulting network objects.

\subsubsection*{Converting wastewater shapefile data to a directed network}
\begin{itemize}
	\item Filter the wastewater data for objects that fall within the perimeter of a \emph{drainage operational area} (DOA) of interest.
	\item Create an initial network as a list of nodes (chambers) and edges (pipes flowing from one chamber to another).
	\item For obvious anomalies in the data (e.g.,~a whole settlement is detatched as the connecting pipe is outside the DOA), create an artificial edge to connect it.
	\item Identify the physical location of the sewage treatment plant associated with that DOA. Label the node closest to that treatment plant as the sink node.
	\item For each node on the network, identify the shortest path to the sink node, by physical distance.
	\item Remove any nodes that are not connected (i.e.~there is no route from that node to the sink), and any edges that are not used on any shortest path. This leaves a single, connected tree, with the sink node as the base.
	\item Associate a directionality to each of the vertices, to move towards the sink (moving from node with degree of separation $n+1$ to node with degree of separation $n$).
\end{itemize}

To associate populations shedding wastewater with each node, we use population at Output Area (OA) level --- census areas containing of order 50 residents:
\begin{itemize}
	\item Identify for each node the OA that it is contained within, and associate that OA to the node. If a node does not fall in an OA (such as pipes  under a body of water), give it a population of zero.
	\item For each OA, evenly distribute the population amongst its associated nodes, maintaining integer population (e.g.~distributing a population of 33 amongst 5 nodes will give populations $(7,7,7,6,6)$).
\end{itemize}

The result is a directed tree where each node has associated with it a physical location, OA, population, and shortest path to the sink.

\clearpage
\section{Additional figures}\label{supp:bonusfigs}

\begin{figure}[H]
	\includegraphics[trim = 0cm 0cm 0cm 0.0cm, clip]{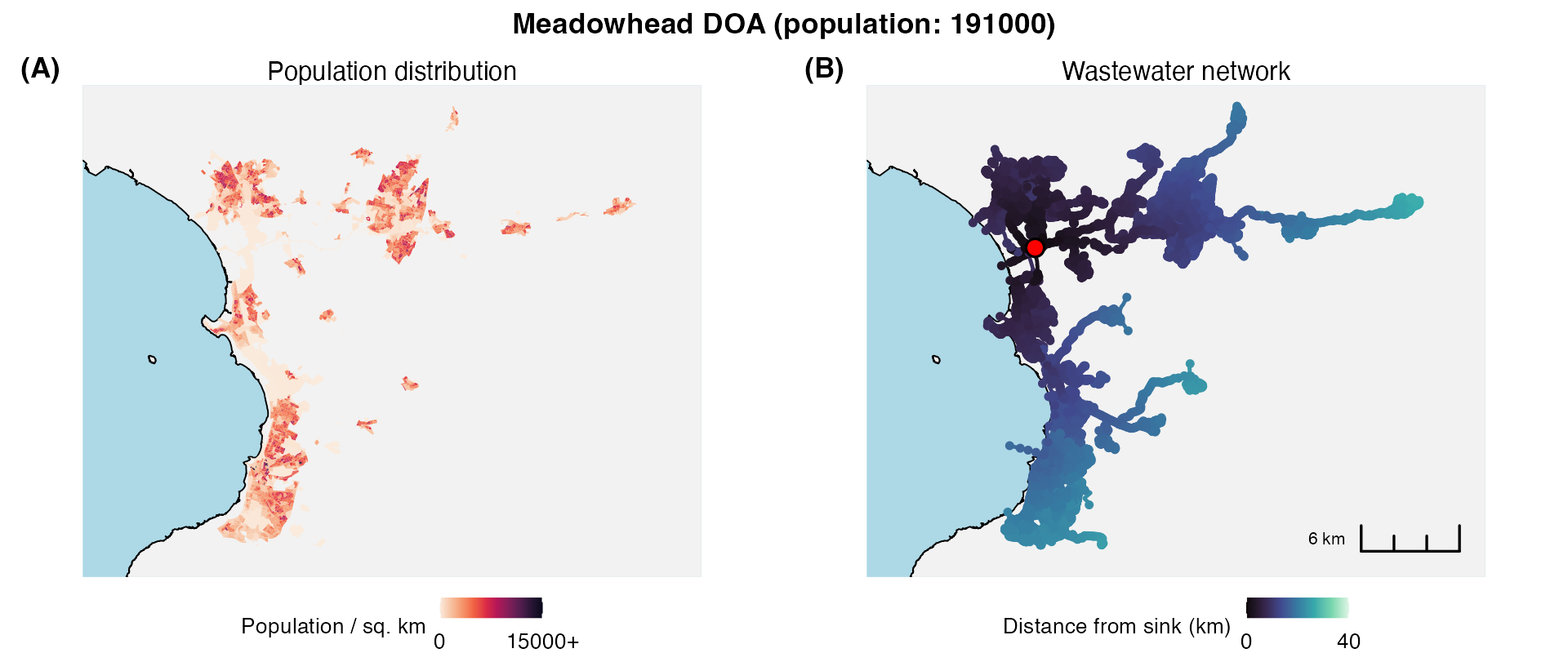}
	\caption{The Meadowhead drainage operational area (DOA). (A) Population distribution. (B) The corresponding wastewater network, where the colour indicates the distance from the treatment plant (marked with a red dot).}
	\label{fig:M_population_and_network}
\end{figure}
\begin{figure}[H]
	\includegraphics[trim = 0cm 0.7cm 0cm 0.0cm, clip]{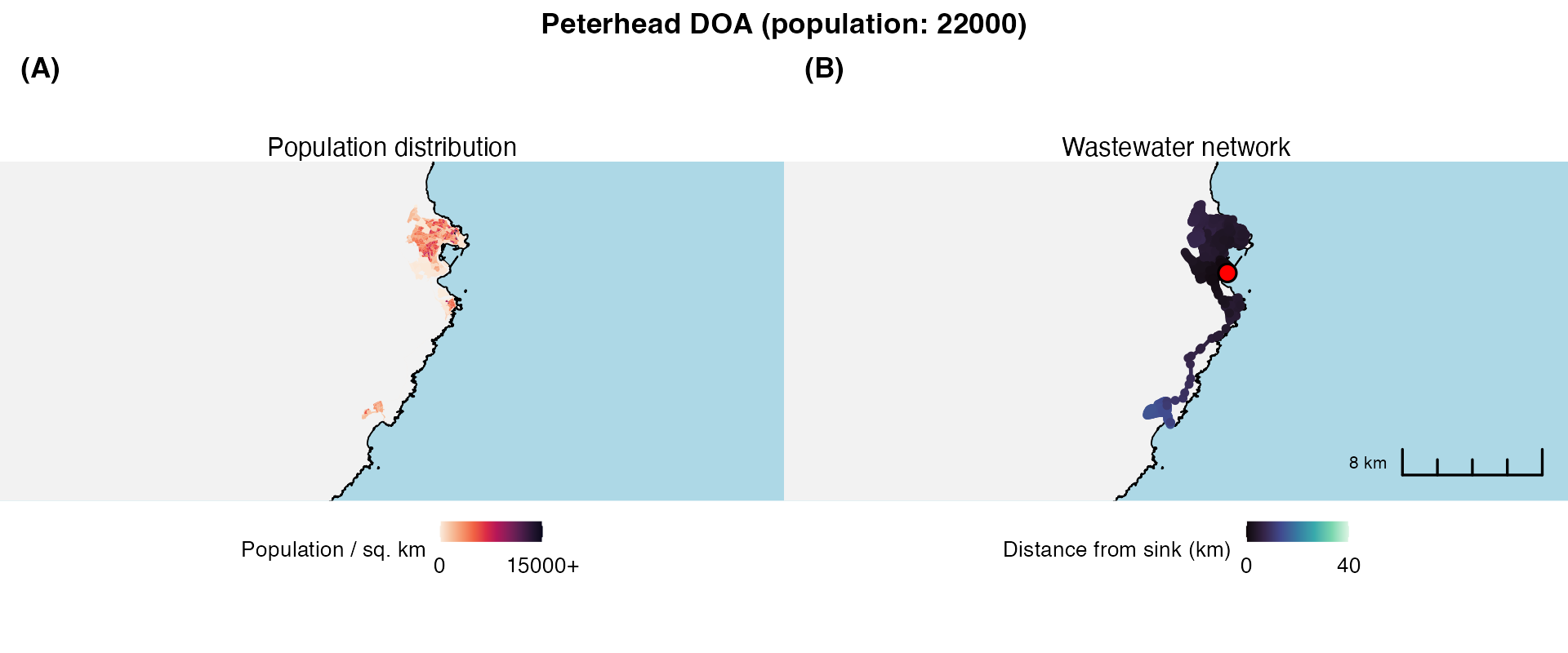}
	\caption{The Peterhead drainage operational area (DOA). (A) Population distribution. (B) The corresponding wastewater network, where the colour indicates the distance from the treatment plant (marked with a red dot).}
	\label{fig:P_population_and_network}
\end{figure}

\clearpage 

\begin{figure}
	\centering
	\includegraphics[trim = 0cm 0.2cm 0cm 0.0cm, clip]{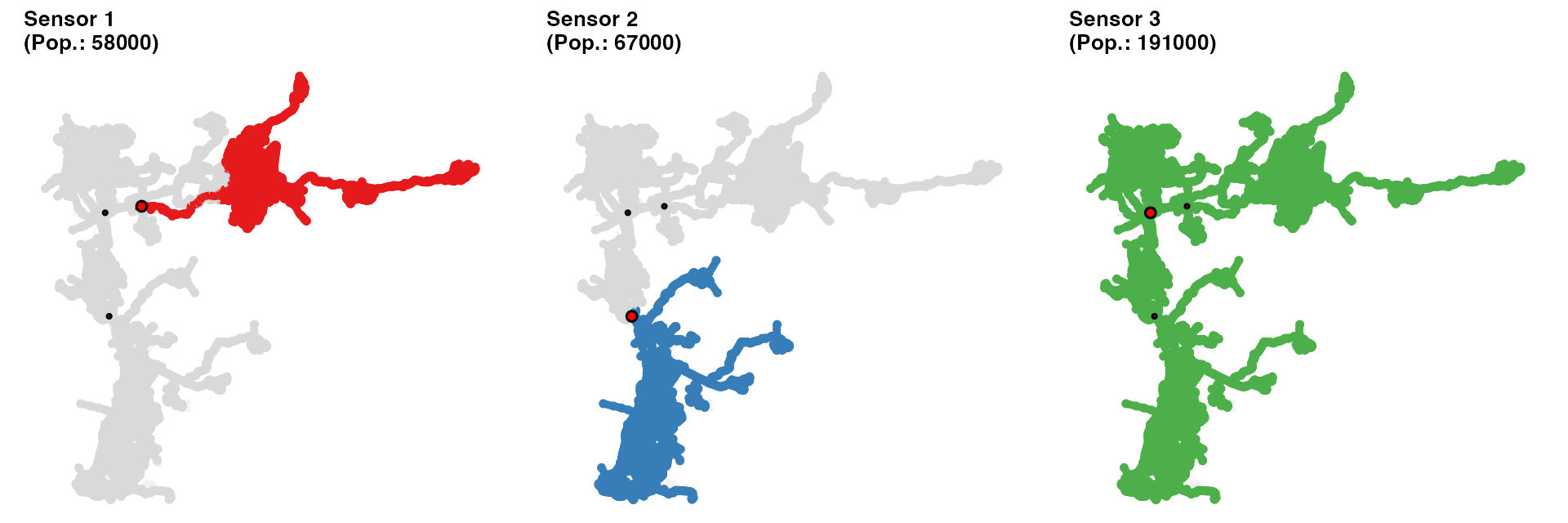}
	\rule{\textwidth}{0.5pt}
	\includegraphics[trim = 0cm 1.0cm 0cm 0.9cm, clip]{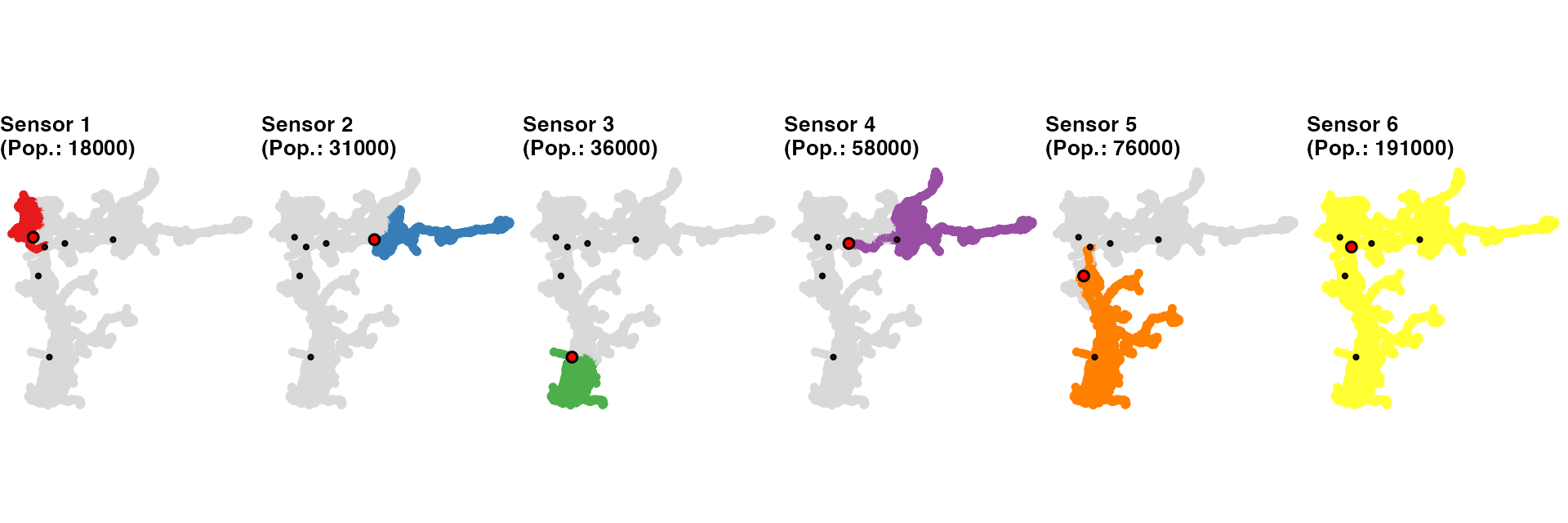}
	\rule{\textwidth}{0.5pt}
	\includegraphics[trim = 0cm 1.3cm 0cm 1.2cm, clip]{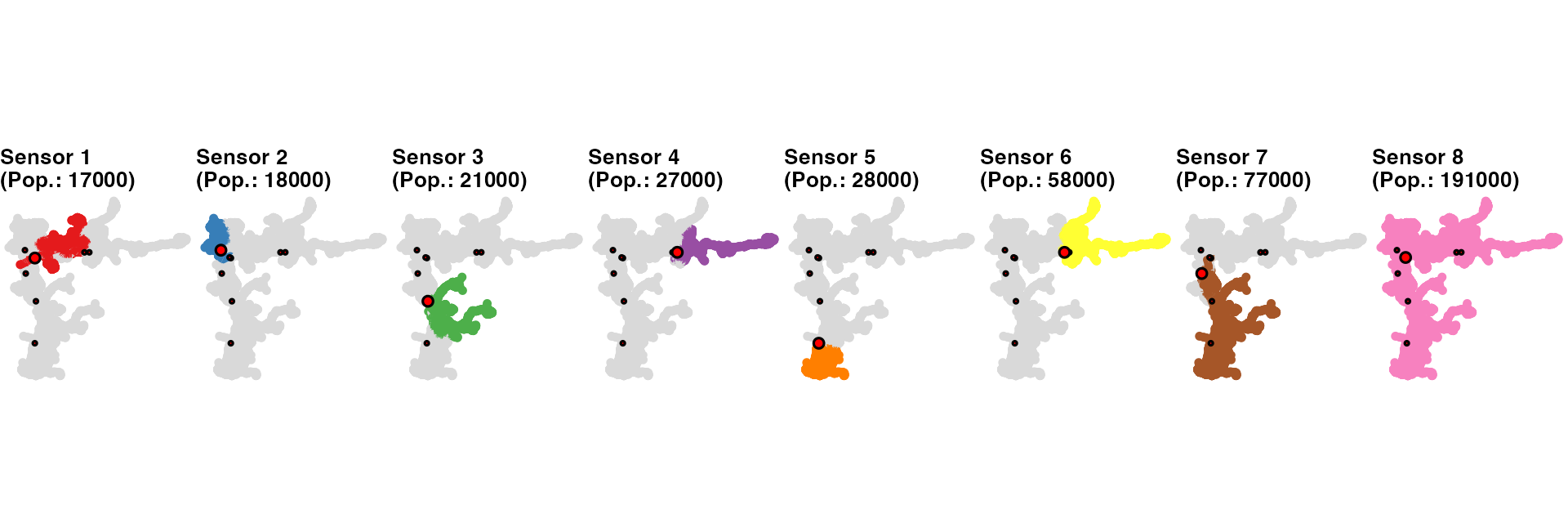}
	\caption{Placement of sensors over the Meadowhead DOA for (top to bottom) 3, 6, and 8 sensors.}
	\label{fig:M_sensor_placement}
\end{figure}

\clearpage

\begin{figure}
	\centering
	\includegraphics[trim = 0cm 0.3cm 0cm 0.0cm, clip]{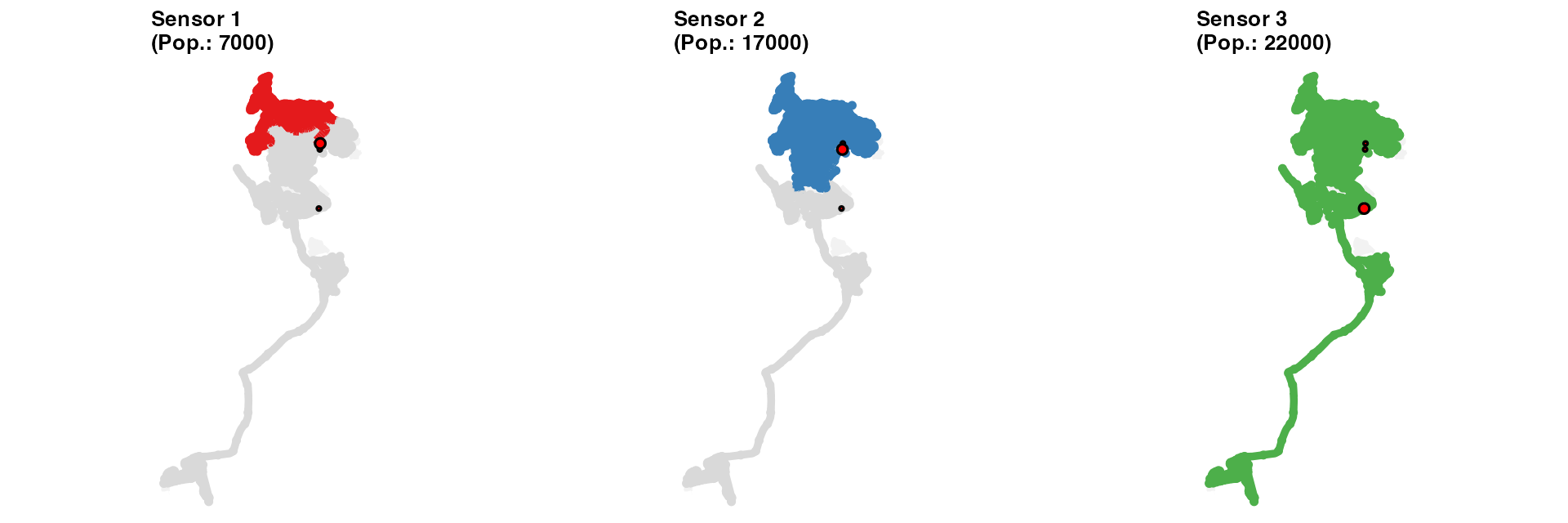}
	\rule{\textwidth}{0.5pt}
	\includegraphics[trim = 0cm 0.3cm 0cm 0cm, clip]{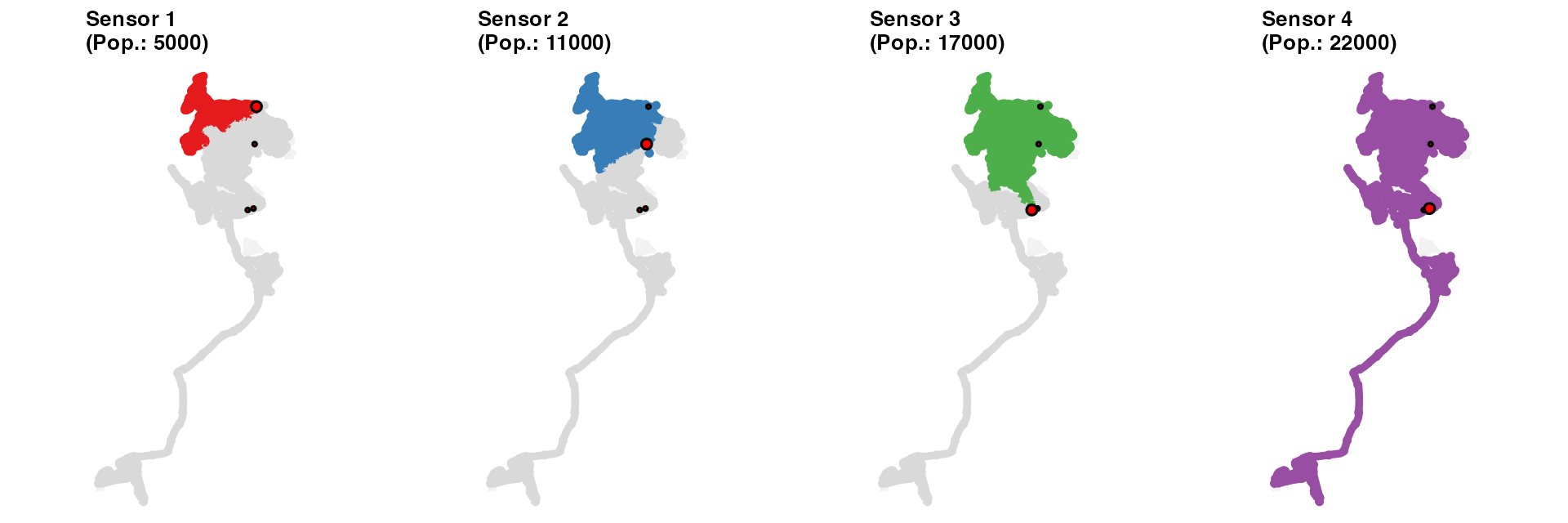}
	\rule{\textwidth}{0.5pt}
	\includegraphics[trim = 0cm 0cm 0cm 0cm, clip]{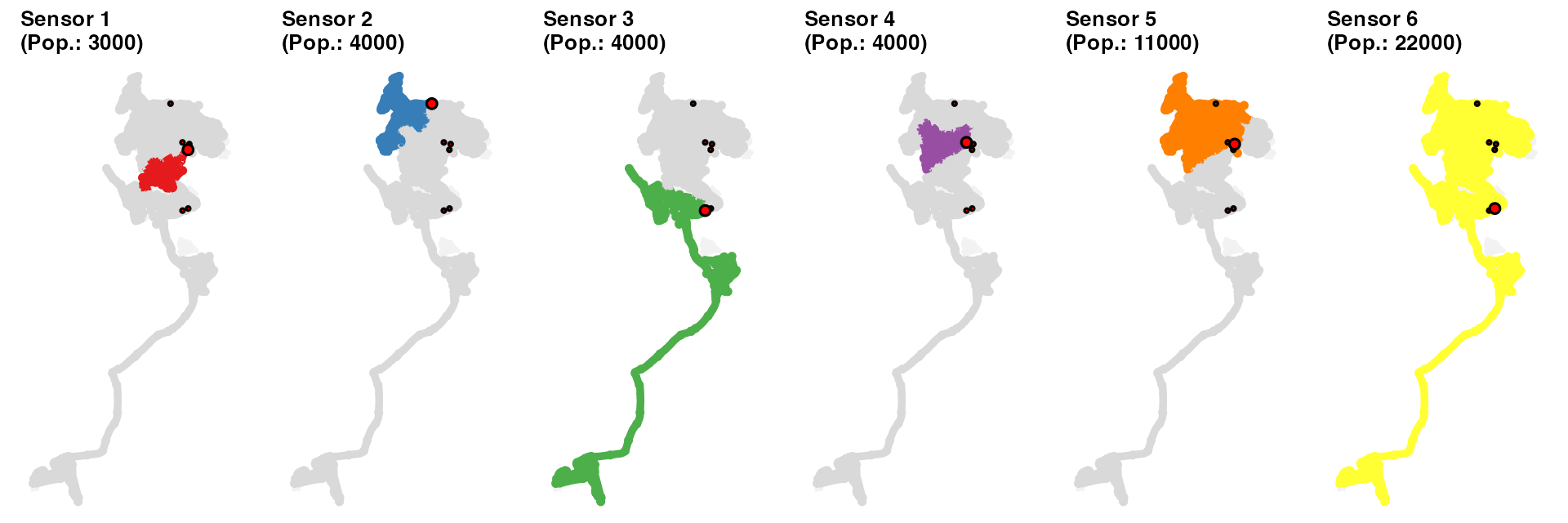}
	\caption{Placement of sensors over the Peterhead DOA for (top to bottom) 3, 4, and 6 sensors.}
	\label{fig:P_sensor_placement}
\end{figure}

\begin{figure}
	\includegraphics[width=0.99\textwidth]{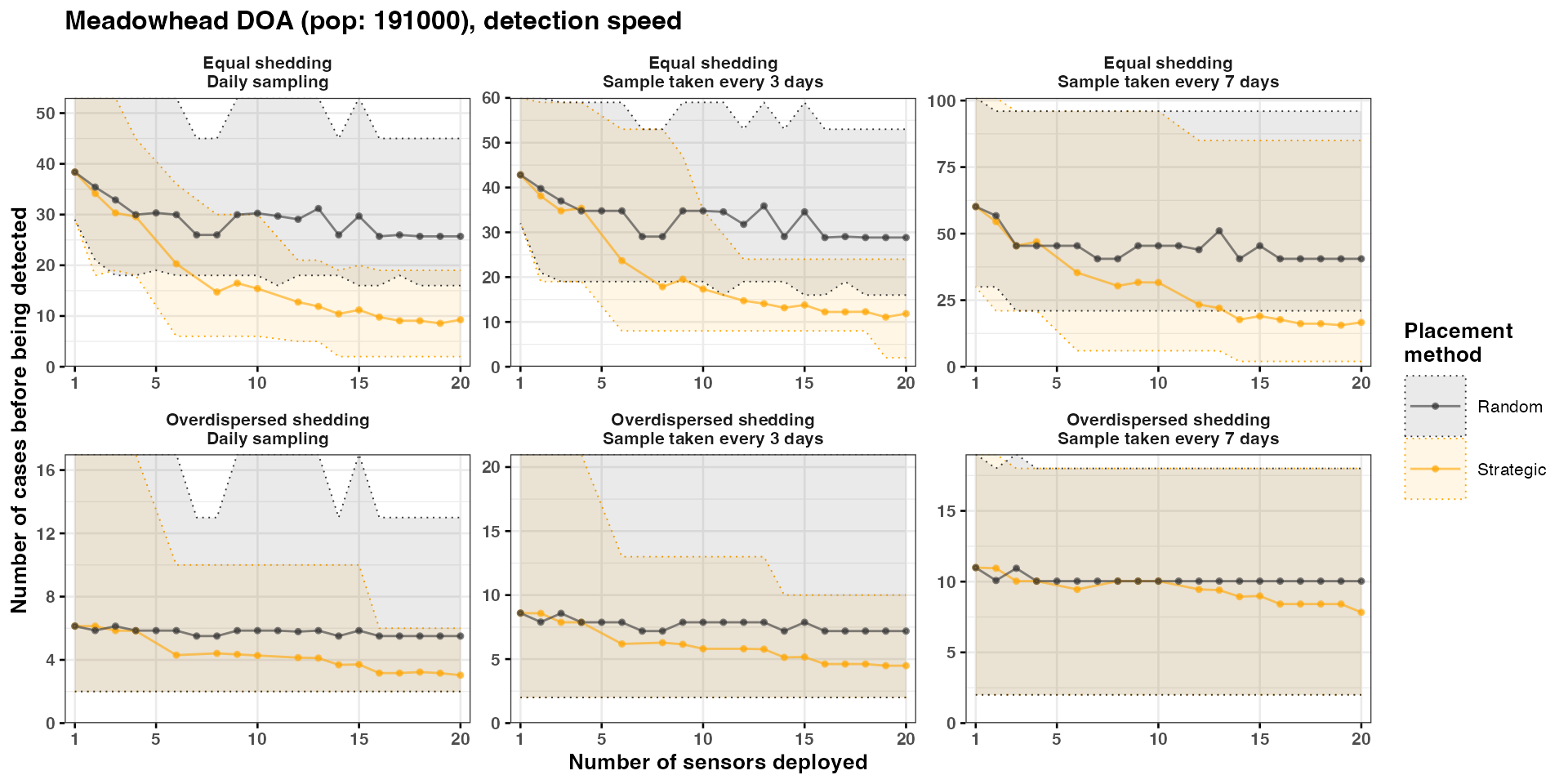}
	\caption{Detection speed for different sensor configurations, in the Meadowhead DOA. Detection speed is quantified by the cumulative disease incidence by the time any sensor detects a non-zero signal. Top: results when shedding is equal (all infected individuals have the same shedding profile with a mean of 1). Bottom: results when shedding is overdispersed (peak shedding is sampled from a lognormal distribution with log mean 1, log standard deviation 3.07).}
	\label{fig:M_first_detection}
\end{figure}

\begin{figure}
	\includegraphics[width=0.99\textwidth]{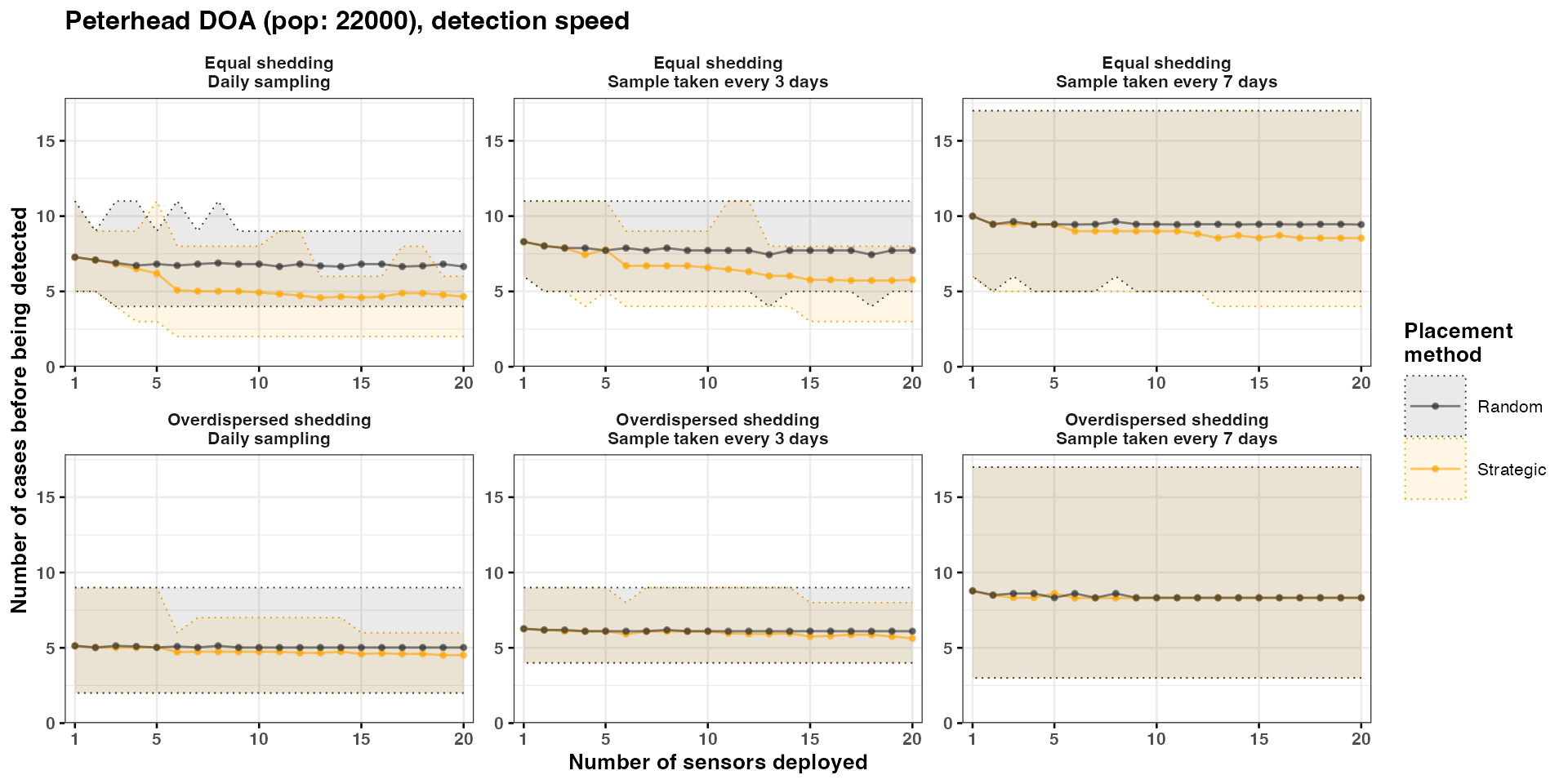}
	\caption{Detection speed for different sensor configurations, in the Peterhead DOA. Detection speed is quantified by the cumulative disease incidence by the time any sensor detects a non-zero signal. Top: results when shedding is equal (all infected individuals have the same shedding profile with a mean of 1). Bottom: results when shedding is overdispersed (peak shedding is sampled from a lognormal distribution with log mean 1, log standard deviation 3.07).}
	\label{fig:P_first_detection}
\end{figure}

\begin{figure}
	\includegraphics[width=0.99\textwidth]{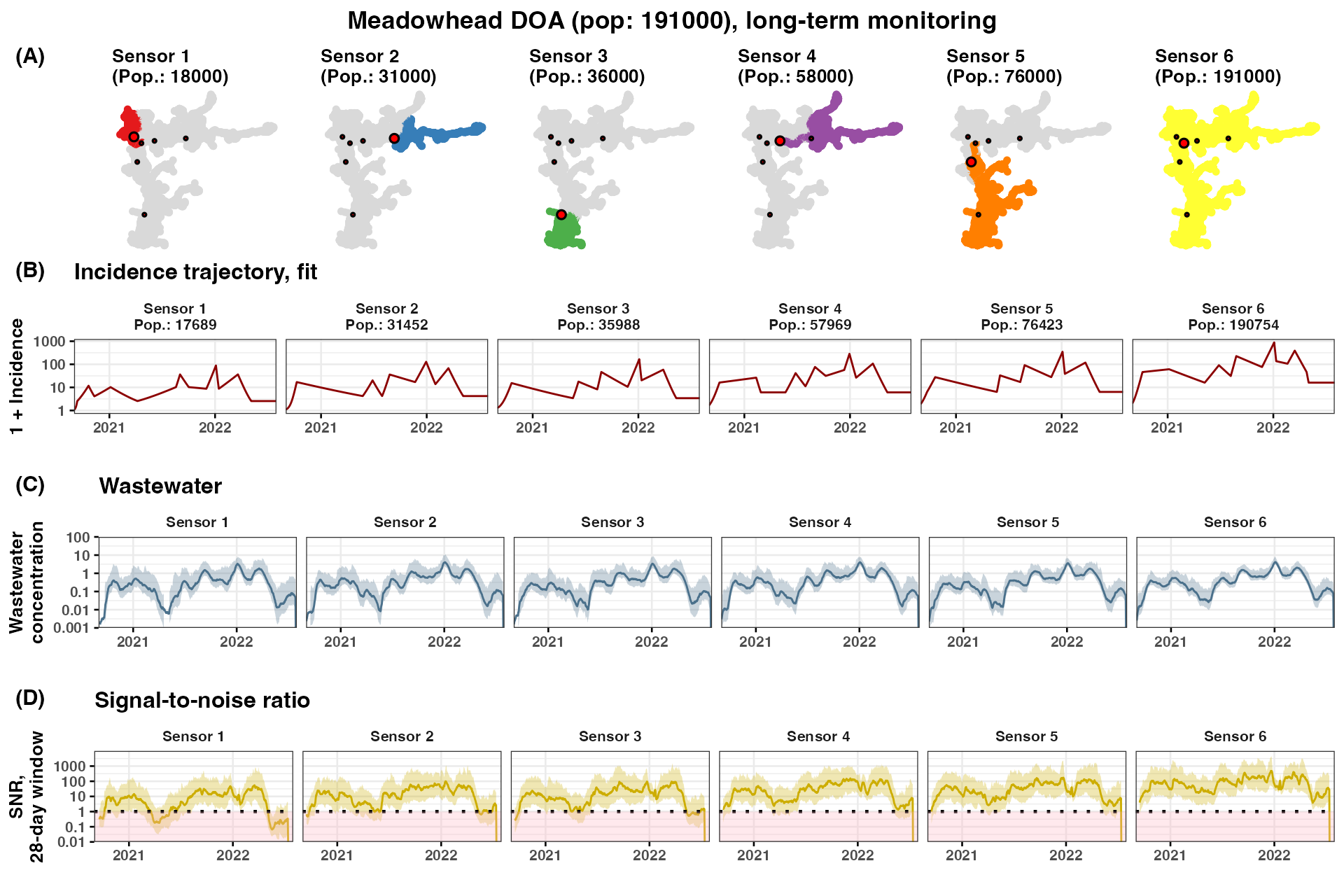}
	\caption{Simulated long-term monitoring of disease in the Meadowhead DOA via wastewater. (A) Placement of an 6-sensor configuration, ordered by population (B) Smoothed COVID-19 incidence trajectory between September 2020 and August 2022. (C) Modelled wastewater signal with the line indicating the median signal, and the filled region the [5\%, 95\%] range of signals. (D) The signal-to-noise ratio of the wastewater signal relative to the smoothed incidence trajectory, over a 28-day rolling window.}
	\label{fig:M_longterm}
\end{figure}

\begin{figure}
	\includegraphics[width=0.99\textwidth]{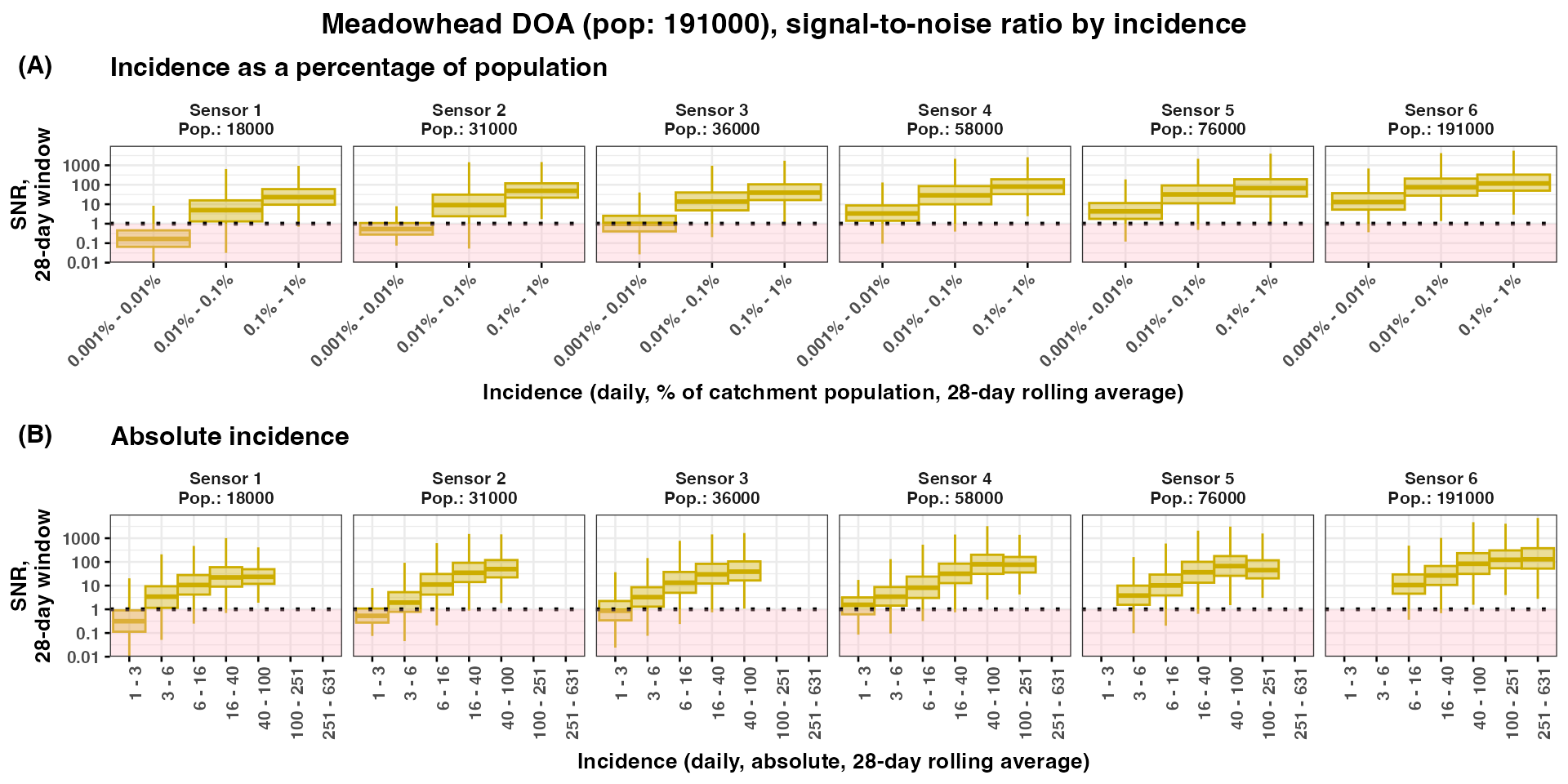}
	\caption{Signal-to-noise ratio statified by disease incidence in the Meadowhead DOA, with 6 sensors. (A) Incidence as a percentage of the catchment population. (B) Incidence as an absolute number of infections.}
	\label{fig:M_longterm_SNR}
\end{figure}

\begin{figure}
	\includegraphics[width=0.99\textwidth]{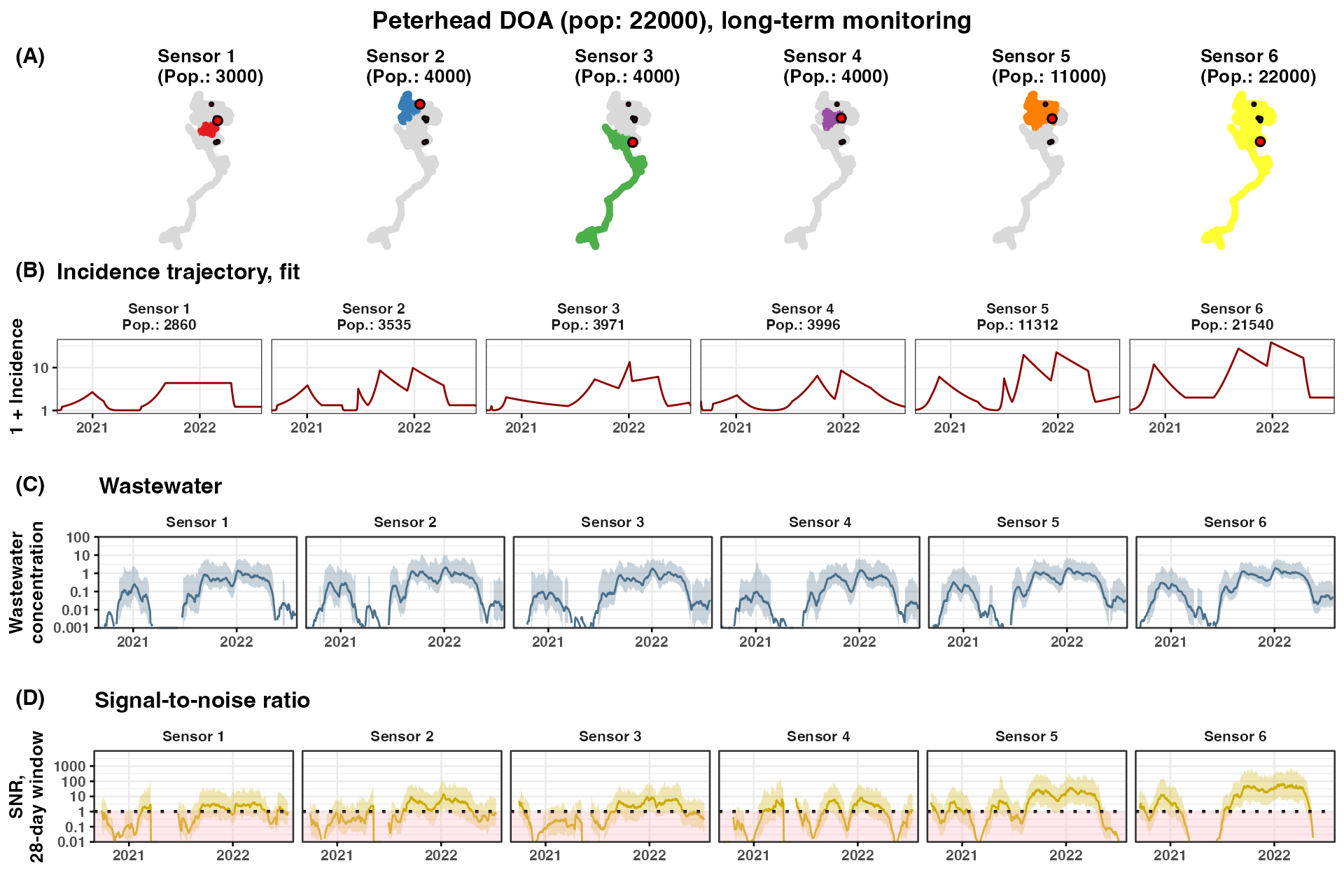}
	\caption{Simulated long-term monitoring of disease in the Peterhead DOA via wastewater. (A) Placement of an 6-sensor configuration, ordered by population (B) Smoothed COVID-19 incidence trajectory between September 2020 and August 2022. (C) Modelled wastewater signal with the line indicating the median signal, and the filled region the [5\%, 95\%] range of signals. (D) The signal-to-noise ratio of the wastewater signal relative to the smoothed incidence trajectory, over a 28-day rolling window.}
	\label{fig:P_longterm}
\end{figure}

\begin{figure}
	\includegraphics[width=0.99\textwidth]{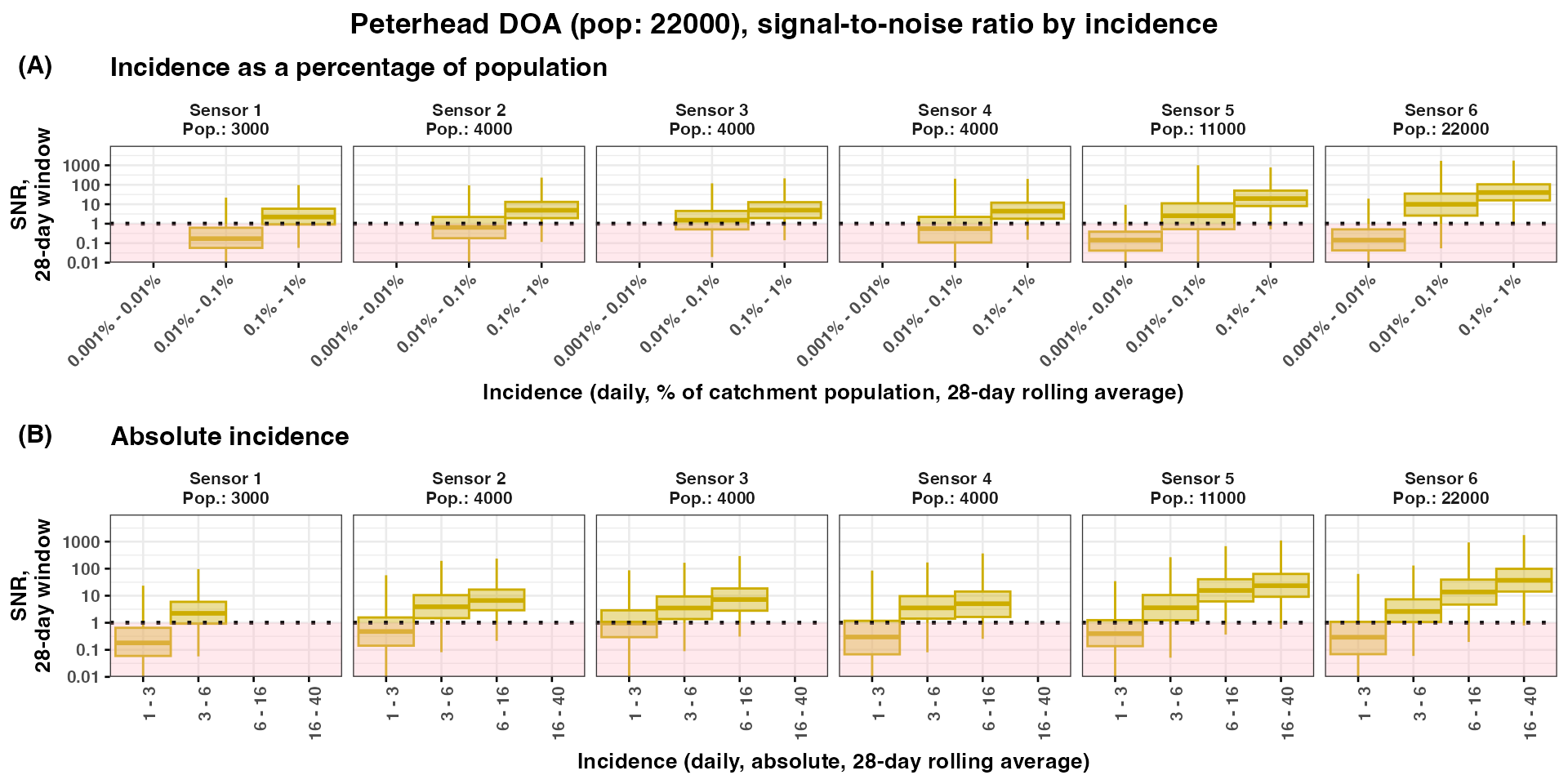}
	\caption{Signal-to-noise ratio statified by disease incidence in the Peterhead DOA, with 6 sensors. (A) Incidence as a percentage of the catchment population. (B) Incidence as an absolute number of infections.}
	\label{fig:P_longterm_SNR}
\end{figure}

\end{document}